\documentclass[leqno,12pt]{article}
\usepackage{graphicx}
\usepackage{epsfig}
\usepackage[latin1]{inputenc}
\usepackage[english]{babel}
\usepackage[T1]{fontenc}
\usepackage{amsfonts,amsmath}
\usepackage{latexsym}
\usepackage{stmaryrd}
\usepackage{amssymb}
\usepackage{theorem}
\usepackage{color}
\input xy
\xyoption{all}

\newlength{\oldparindent}
\title{\Large \bf Information Asymmetry in Pricing of Credit Derivatives}
\author{\textsc{Caroline HILLAIRET\thanks{CMAP Ecole Polytechnique, Email: caroline.hillairet@polytechnique.edu}\quad  Ying JIAO\thanks{LPMA Universit\'e Paris 7, Email: jiao@math.jussieu.fr}}\\
}

\newtheorem{definition}{Definition }[section]
\newtheorem{proposition}[definition]%
{Proposition }
\newtheorem{lemme}[definition]%
{Lemma }\newtheorem{theoreme}[definition]%
{Theorem }
{Corollary }
{Remark }
\newtheorem{hypothese}[definition]%
{Assumption}

\theorembodyfont{\upshape}
\newtheorem{ex}[definition]%
{Example}

\newcommand{\esp}{E} 

\newcommand{\indic}{1\!\!\!1} 
\newcommand{\proba}{\mathbb{P}}

\newcommand{\Q}{\mathbb{Q}}
\newcommand{\R}{\mathbb{R}}
\newcommand{\D}{\mathcal{D}}
\newcommand{\F}{\mathcal{F}}

\newcommand{\G}{\mathcal{G}}

\theorembodyfont{\upshape\it}
\newtheorem{Thm}{\bf Theorem}[section]
\newtheorem{Pro}[Thm]{\bf Proposition}

\theorembodyfont{\upshape}

\newtheorem{Rem}[Thm]{Remark}

\newcommand {\bproof} {\noindent {\sc Proof:  }}
\newcommand {\finproof} {\hfill $\Box$ \vskip 5 pt }
\def\rmk{\noindent \textit{Remark: }}

\begin{document}

\maketitle

\numberwithin{equation}{section}


\begin{abstract}
We study the pricing of credit derivatives with asymmetric information. 
The managers have complete information on
the value process of the firm and on the default threshold, while the investors on
the market have only partial observations, especially about the default threshold. 
Different information
structures are distinguished using the framework of enlargement of filtrations. We specify risk neutral
probabilities and we evaluate default sensitive contingent claims
in these cases.

\end{abstract}


\newpage

\section{Introduction}
\addcontentsline{toc}{section}{Introduction}
The modelling of a default event is an important subject from both economic and financial point of view. There exist a large literature on this issue and mainly two modelling approaches: the structural one and the reduced-form one. In the structural approach, where the original idea goes back to the pioneer paper of Merton \cite{merton74}, the default is triggered when a fundamental process $X$ of the firm passes below a threshold level $L$. The fundamental process may represent the asset value or the total cash flow of the firm where
the debt value of the firm can also be taken into consideration. This provides a convincing economic interpretation for this approach. The default threshold $L$ is in general  supposed to be constant or deterministic. Its level is chosen by the managers of the firm according to some criterions --- maximizing the equity value for example as in \cite{leland}.

For an agent on the financial market, the vision on the default is quite different: on one hand, he possesses merely a limited information of the basic data (the process X for example) of the firm; on the other hand, to deal with financial products written on the firm, he needs to update his estimations of the default probability in a dynamic manner. This leads to the reduced-form approach for default modeling where the default arrives in a more ``surprising'' way and the model parameters can be daily calibrated by using the market data such as the CDS spreads.

The default time constructed in the classical structural approach is a stopping time with respect to the filtration $\mathbb F$ generated by the fundamental process. The intensity of such predictable stopping times does not exist. In the credit risk literature, it is also interpreted by the fact that the default intensity (or the credit spread) tends to zero when the time to maturity decreases to zero (we shall make precise the meanings of these two intensities later on). The links between the structural and the intensity approaches have been  investigated in the literature. If the default threshold $L$ is a random variable instead of constant or deterministic, then the default time admits the intensity. One important example is the well known Cox process model introduced in \cite{lando} where $L$ is supposed to be an exponentially distributed random variable independent with $\mathbb F$ (see also \cite{ElK}).  Another class of models  is the incomplete information models (e.g. \cite{duffielando, C-DGH, jeanblancvalchev, CGJ, CJPY}) where the agent only has a partial observation of the fundamental process $X$ and thus his available information is represented by some subfiltration of $\mathbb F$. The intensity can then be deduced for the subfiltration.

In this paper, we are interested in the impact of information accessibility of
an agent on the pricing of credit derivatives. In particular, we aim to study
the information concerning the default threshold $L$  in addition to the
partial observation of the process $X$. This case has been studied in
\cite{GG08} where investors anticipate the distribution of $L$ (following for example the Beta distribution) whose parameters are
calibrated through market data.  Our approach is different and is related to the insider's information problems. Indeed, when the managers make decisions on whether the firm will default or not, he has supplementary information on the default threshold $L$ compared to an ordinary investor on the market. Facing the financial crisis, this study is also motivated by some recent ``technical default events'', where the bankruptcy occurs although the firm is still capable to repay its debts.

We present our model in the standard setting. Let $(\Omega, \mathcal{A}, \mathbb{P})$  be a probability space which represents the financial market. We consider a firm and model its default time as the first time that a { continuous time process  $(X_t)_{t \geq 0}$} reaches some default barrier  $L$, i.e.,
\begin{equation}\label{equ: default model}
\tau=\inf\{t: X_t \leq L\} \quad \text{ where } X_0>L
\end{equation}
with the convention that $\inf \emptyset= + \infty$. Denote by $\mathbb F=(\F_t)_{t\geq 0}$ the filtration generated by the process
$X$, i.e., $\F_t=\sigma(X_s,s\leq t)\vee \mathcal N$ satisfying the usual conditions where $\mathcal{N}$ denotes the $\mathbb{P}$ null sets. Such construction of a default time adapts to both the structural approach and  the reduced form approach of the default modelling,  according to the specification of the process $X$ and the threshold $L$.

In the structural approach models, $L$ is a constant or a deterministic function $L(t)$, then $\tau$ defined in \eqref{equ: default model} is an $\mathbb F$-stopping time as in the classical first passage models.
In the reduced-form approach, the default barrier $L$ is unknown and is described as a random variable in $\mathcal A$. We introduce the decreasing process $X^*$ defined as
$$X^*_t=\inf\{X_s, \,s\leq t\}.$$
Then \eqref{equ: default model} can be rewritten as
\begin{equation}\label{equ:tau defined in intensity approch}
\tau=\inf\{t: X^*_t=L\}.\end{equation}
This formulation gives a general reduced-form model of default (see \cite{ElK}). In particular, when the barrier $L$ is supposed to be independent of $\F_{\infty}$,  then $$\proba(\tau>t|\F_{\infty})=\proba(X^*_t > L|\F_{\infty})=F_L(X^*_t),$$ where $F_L$ denotes the distribution function of $L$. Note that the (H)-hypothesis is satisfied in this case, that is, $\proba(\tau>t|\F_{\infty})=\proba(\tau>t|\F_t)$. We may also recover the Cox-process model using a similar construction.

In most papers  concerning the information-based credit models, the process $X$ is partially observed, making an impact on the conditional default probabilities and on the credit spreads. In this paper, we let $L$ to be a random variable and  take into consideration the information on $L$. Such information modelling is closely related to the enlargement of filtrations theory.  Generally speaking, the information of a manager is represented by the initial enlargement of the filtration $(\F_t)_{t\geq 0}$ and the  information of an investor is modelled by the progressive enlargement of $(\F_t)_{t\geq 0}$ or of some of its subfiltration.   We shall also consider the case of an insider who may have some extra knowledge on $L$ compared to an investor and whose knowledge is however perturbed compared to the manager.

The rest of this paper is organized as follows. In Section 2, we
introduce the pricing problem and the different information structures for various agents on
the market,  notably the information on the
default barrier $L$. We shall distinguish the role of the manager, the investor and the insider, who have different level of information on $L$.  
In successively Sections 3, 4 and 5, 
we  make precise  the mathematical hypothesis for these cases, using the
languages of enlargement of filtrations.  We also discuss the
risk-neutral probabilities in each case for further pricing
purposes. In order to distinguish the
impact of the different filtrations from the impact of the
different pricing probabilities, we first give the price of a
contingent claim under the historical probability measure
$\mathbb{P}$ for each information in Section \ref{sec:full},
\ref{sec:prog}, \ref{sec:noisy}, the calculus under the
corresponding  pricing (or "risk-neutral") probability   being
done in the last section. Finally, we end the last section with numerical illustrations.

\section {Pricing framework and information structures}
\label{Sec:general}
On the financial market, the available information for each agent is various. There exists in general information asymmetry between different market investors, and moreover between the managers of a firm and the investors.  In particular, the managers may have information on whether the firm will default or not, or when the default may happen. The pricing of credit-sensitive derivative depends strongly on the information flow of the agent. We begin by  introducing the general pricing principle and then we precise different information.

\subsection{General pricing principle}

We fix in the sequel a probability space $(\Omega,\mathcal
A,\mathbb P)$ and a filtration $\mathbb F=(\mathcal F_t)_{t\geq
0}$ of $\mathcal A$, representing the default-free information.
Let $\tau$ be a strictly positive and finite random time on
$(\Omega,\mathcal A,\mathbb P)$, modelling the default time. The
information flow of the agent is described by a filtration
$\mathbb H=(\mathcal H_t)_{t\ge 0}$ such that $\tau$ is an
$\mathbb H$-stopping time, that is, all agents observe at time $t$
whether the default has occurred or not. Without loss of generality, we assume that all the filtrations we consider satisfy the usual conditions of completeness and right-continuity. 

We describe a general credit-sensitive derivative claim
of maturity $T$ as in \cite{BR},
by a triplet $(C,G,Z)$
where $C$ is an $\F_T$-measurable random variable
representing the payment at the maturity $T$ if no default occurs before the maturity, $G$ is an
$\mathbb F$-adapted, continuous process of finite
variation with $G_0=0$ and represents the dividend
payment, $Z$ is an $\mathbb F$-predictable process and
represents the recovery payment at the default time
$\tau$. 

 The triplet for a CDS, viewed by a
protection buyer, satisfy $C=0$, $G_t=-\kappa t$ and $Z=1-\alpha$
where $\kappa$ is the spread of CDS and $\alpha$ is the recovery rate of
the underlying name. The triplet for a defaultable zero-coupon satisfy $C=1$, $G=0$ and $Z=1-\alpha$.

The  value process of the claim at time $t<\tau\wedge T$ is given by
\begin{equation}\label{Equ:value process}
V_t=R_t\esp_{\mathbb Q}\Big[CR_T^{-1}\indic_{\{\tau > T\}}+ \int_t^T
\indic_{\{\tau>u\}}R_u^{-1}dG_u+Z_{\tau}\indic_{\{\tau \leq  T\}}R_{\tau}^{-1}
\,\Big|\,\mathcal H_t\Big]\end{equation} where $\mathbb Q$ denotes the
pricing  probability measure which we shall precise later, and $R$ is the discount factor process. We note that both the filtration and the pricing probability depend on the information level of the agent.

In the credit risk analysis, one often tries to establish a
relationship between the market filtration and the default-free
one. The main advantage is that the default-free filtration is
often supposed to have nice regularity conditions, while the
global market filtration which contains the default information is
often difficult to work with directly. Indeed, due to the default
information, the processes adapted to the global filtration have
in general a jump at the default time (except in the structural
approach) and this makes it difficult to propose explicit models
in this filtration. In our model with insider's information, we
need to make precise the filtration $\mathbb H=(\mathcal
H_t)_{t\geq 0}$ in \eqref{Equ:value process} for different types
of agents. Our objective, similar as mentioned above, is to
establish a pricing formula with respect to the default-free
filtration in each case.

\subsection{Information structures}
We  now describe the different information flows and the
corresponding filtration $\mathbb H$ for different agents on the
market. Recall that the default time is modelled by
\[\tau=\inf\{t\,:\,X_t^*=L\},\]
where $L$ is a random variable and $X^*$ is the infimum process of an $\mathbb F$-adapted process $X$.
We assume that $L$ is chosen by the managers of the firm who hence have the  total knowledge on $L$. The information of $X^*_t$ is contained in the $\sigma$-algebra $\F_t$. However, the process $X^*$ can not give us full information on $\F_t$.

$\bullet$ Manager's information.\\
The manager has complete information on $X$ and on $L$. The
filtration of the manager's information, denoted by $\mathbb
G^M=(\mathcal G_t^M)_{t\geq 0}$, is then
$$\mathcal{G}^{M}_t :=  \mathcal{F}_t  \vee \sigma(L).$$
Note that $\mathbb G^M$ is in fact the initial enlargement of the
filtration $\mathbb F$ with respect to $L$ and we call it the full information on $L$.
It is obvious that $\tau$ is a $\mathbb G^M$-stopping time.  We
shall precise some technical hypothesis in the next section.

$\bullet$ Investor's information.\\
In the credit risk literature, the accessible information on the
market is often modelled by the progressive enlargement $\mathbb
G=(\G_t)_{t\geq 0}$ of $\mathbb F$. More precisely, let $\mathbb
D=(\D_t)_{t\geq 0}$ be the minimal filtration which makes $\tau$ a
$\mathbb D$-stopping time, i.e. $\D_t=\D_{t+}^0$ with
$\D_t^0=\sigma(\tau\wedge t)$, then $$\G_t=\F_t\vee\D_t.$$ In our
model \eqref{equ:tau defined in intensity approch}, this is
interpreted as $\G_t=\F_t\vee\sigma (\{L\leq X^*_t\})$ and we call
this information the progressive (enlargement) information  on $L$.
Together with the 
information flow of the filtration $(\F_t)_{t\geq 0}$, an investor who observes the filtration $(\G_t)_{t\geq 0}$ knows at time $t$  whether or not the default
has occurred up to $t$ and the default time $\tau$ once it occurs. We see that the manager's information
$\G^M_t$ is larger than $\G_t$.

$\bullet$ Investor's incomplete information.\\
In many incomplete information credit risk models, the process $X$ driving the
default risk is not totally observable for the investors. In this paper, we
will only consider the example of a delayed information on $X$ :  the information of such an investor is described by a progressive enlargement $\mathbb G^D=(\mathcal G_t^D)_{t\ge 0}$ of a delayed filtration of $\mathbb F$, where
\[\mathcal G_t^D:=\mathcal F_{t-\delta(t)}\vee\D_t,\]
and $\delta(t)$ being a function valued in $[0,t]$ such that
$t-\delta(t)$ is increasing. The above formulation covers the
constant delay time model where $\delta(t)=\delta$ (see
\cite{C-DGH}, \cite{GJZ}) and the discrete observation model where
$\delta(t)=t-t_i^{(m)}$ and $t_i^{(m)}\leq t<t_{i+1}^{(m)}$,
$0=t_0^{(m)}<t_1^{(m)}<\cdots<t_m^{(m)}=T$ being the discrete
dates on which the $(\F_t)_{t\geq 0}$ information may be renewed
(for example, the release dates of the accounting reports of the
firm, see \cite{duffielando}, \cite{jeanblancvalchev}).

$\bullet$ Insider's information.\\
Finally, we shall consider the insiders who have as supplementary
information a partial observation on $L$ compared to the
investor's information $\G_t$. Namely, the agent has the knowledge on a noisy default threshold:
$(L_t)_{t\ge 0}$, $L_s=f(L, \epsilon_s)$ with{ $\epsilon$ being an
independent noise perturbing the information on $L$}. The
corresponding information flow is then modelled by $\mathbb
G^I=(\mathcal G_t^I)_{t\ge 0}$ where
\[\mathcal G^{I}_t:=\mathcal F_t\vee\sigma(L_s,\,s\le t)\vee\D_t.\]
Notice that $\mathcal G^{I}_t=\G_t\vee\sigma(L_s,\,s\le t)$. We
call this information the ``noisy full information'' on $L$. It is
a successive enlargement of $\F_t$, firstly by the noised
information of the default threshold and then by the default occurrence
information.

For the different types of information described above, we observe
that the following relations hold:
\[\mathbb G^M\supset\mathbb G^I \supset\mathbb G\supset\mathbb G^D.\]
They correspond to the pricing filtration $\mathbb H$ in
\eqref{Equ:value process} for different agents on the market. We
shall concentrate on the pricing problem with the above
filtrations and we begin by making precise the mathematical
hypothesis on these types of information on $L$, {with which we
introduce the risk-neutral probabilities $\Q$ in each case.}

\section{Full information}\label{sec:full}
In this section, we work with the manager information flow $\mathbb G^M=\mathbb F\vee\sigma(L)$, which is an initial enlargement of the filtration $\mathbb F$. Recall that the default barrier is fixed  at date 0 by the manager as the realization of a random variable $L$.
We assume in addition that the filtration $\mathbb F$ is generated by a Brownian motion $B$.

\subsection{Initial enlargement of filtration}\label{Subsec:jacod}

In the theory of initial enlargement of filtration, it is standard to work under the following density hypothesis due to Jacod \cite{Jac, Jac2}.

\begin{hypothese}\label{hyp1}
We assume that $L$ is an  $\mathcal{A}$-measurable random variable
with values in  $\mathbb R$, which satisfies the assumption :
$$\mathbb{P}(L\in \cdot |\mathcal{F}_{t})(\omega) \sim\mathbb{P}(L\in \cdot ), \quad \forall t \geq 0,\,   \mathbb{P}-a.s..$$
\end{hypothese}
\rmk Jacod has shown that,  if Assumption \ref{hyp1} is fulfilled, then any $\mathbb F$-local martingale is a $\mathbb G^M$-semimartingale.

We denote by $P^L _t(\omega,dx)$ a regular version of the
conditional law of $L$ given $\mathcal{F}_t $  and by {$P^{{L}}$}
the law of $L$ (under the probability $\mathbb P$). According to
\cite{Jac2}, there exists a measurable version of the conditional
density
\begin{equation}\label{pt}p_t(x)(\omega)=\frac{dP^L _t}{dP^{L}}(\omega,x)\end{equation}  which is an
$(\mathbb F,\mathbb{P})$-martingale and hence can be written as
$$p_t(x)=p_0(x)+ \int_0^t \beta_s(x) d B_s, \quad\forall x\in\R$$
for some $\mathbb F$-predictable process $(\beta_t(x))_{t\ge 0}$.
Moreover, the fact that $P_t^{L}$ is equivalent to $P^L$ implies
that $\mathbb{P}$-almost surely $p_t(L)>0$. Let us introduce the
$\mathbb F$-predictable process $\rho^M$ where
$\rho_t^M(x)={\beta_t(x)}/{p_t(x)}$, the density process $p_t(L)$
satisfies the following stochastic differential equation
\[dp_t(L)= p_t(L) \rho_t^M(L) dB_t.\]
Note that $ (\widetilde{B}^M _t:= B_t- \int_0^t \rho_s^M(L)
ds,t\geq 0)$ is a $(\mathbb G^{M},\mathbb{P})$-Brownian motion.

It is proved in \cite{Pon} that Assumption \ref{hyp1} is satisfied
if and only if there exists a probability measure equivalent to
$\mathbb{P}$ and  under which $\mathcal{F}_{\infty}:=\cup_{t\ge
0}\mathcal F_t$ and $\sigma(L)$  are independent. The probability
$\mathbb{P}^{L}$ defined by the density process
$$\esp_{\mathbb{P}^{L}}  \Big[ \frac{ \mathrm{d}\mathbb{P}}{\mathrm{d}
\mathbb{P}^{L}}\big|\,\mathcal{G}^M_t\Big]=p_t(L)$$ is the only
one that is identical to   $\mathbb{P}$ on $\mathcal{F}_{\infty}$.

We introduce the process $Y^M$ by
\begin{equation}\label{defYk1}Y^{M}=\mathcal{E}  \bigg(  - \int_0^{\cdot} \rho^M_s(L) d\widetilde{B}^M_s\bigg),\end{equation}
where $\mathcal E$ denotes the Dol\'eans-Dade exponential. We
assume in addition that $Y^M$ is a $(\mathbb G^{M},\mathbb{P})$
martingale. A straightforward computation yields
$d((Y_t^{M})^{-1})=(Y_t^{M})^{-1}{\rho_t^M(L)}dB_t.$ Thus, $Y_t^M=\frac{1}{p_t(L)}$,
that is, $Y^M_t$ is the Radon-Nikodym density of the change of
probability $\proba^L$ with respect to $\proba$ on $\G^M_t$. The
process $Y^M$ is important in the study of risk-neutral
probabilities on $\mathbb G^M$. Indeed, let $\phi$ be the price
process of a default-free financial instrument. It is an $\mathbb
F$-adapted process which is an $\mathbb F$-local martingale under
certain $\mathbb F$ risk-neutral probability $\mathbb Q$ (which is
equivalent to $\mathbb P$). In general $\phi$ is not an $(\mathbb
G^M,\mathbb Q)$-local martingale. However, if we define a new
probability measure $\mathbb Q^M$ by
\[\mathrm{d}\mathbb Q^M=Y_t^{M}\mathrm{d}\mathbb Q\quad\text{on $\mathcal G^{M}_t$},\]
then any $(\mathbb F,\mathbb Q)$-local martingale is an $(\mathbb
G^M,\mathbb Q^M)$-local martingale. In particular, $B$ is a
$(\mathbb G^M,\mathbb Q^M)$-Brownian motion. Moreover, one has the
following martingale representation property by \cite{Am1}: if $A$
is a $(\mathbb G^M,\mathbb Q^M)$-local martingale, then there
exists $ \psi \in L^1_{loc}({B},\mathbb{G}^{M},\mathbb{Q}^M )$
such that
\[ {A}_t= {A}_0+\int_0^t  \psi_s dB_s.\]
This shows that the market is complete for the manager.

\subsection{Pricing with full information}

We consider now  the pricing problem with the manager's
information flow $\mathbb H=\mathbb G^M$ and we assume Assumption
\ref{hyp1}. In order to distinguish the impact of  different
filtrations and the impact of different pricing measures, we first
assume that the pricing probability is $\mathbb P$ for all agents.
The result under $\Q^M$, the risk-neutral probability for the
manager, is computed  in Section
\ref{riskneutral} by a change of probability measure.

Our objective is to establish the pricing formula for the manager
with respect to the default-free filtration $\mathbb F$. We begin
by giving the following useful result.

\begin{Pro}\label{Propprobfull}
For any $\theta\ge t$ and any positive $\mathcal F_\theta\otimes\mathcal B(\mathbb R)$-measurable function $\phi_\theta(\cdot)$, one has
\begin{equation}\label{Equ:probfull}
E_{\mathbb{P}} [\phi_{\theta}(L)\indic_{\{\tau>\theta\}} \,|\, \mathcal{G}^M_t]= \frac{1}{p_t(L)} E_{\mathbb{P}} [\phi_{\theta}(x)p_{\theta}(x) \indic_{\{ X^*_\theta >
  x\}} \,|\, \mathcal{F}_t ]_{x=L} \end{equation}
where $p_t(x)$ is defined in \eqref{pt}.
\end{Pro}
\bproof Let $\mathbb P^L$ be the equivalent probability measure of $\mathbb P$ of density $p_t(L)^{-1}$ on $\mathcal G_t^M$.
By using  the facts  that $\mathcal{F}_{\theta}$ and  $\sigma(L)$  are
independent under  $ \mathbb{P}^{L}$ and that
$\mathbb{P}^{L}$   is identical to   $\mathbb{P}$ on
$\mathcal{F}_{\infty}$,  we have

\begin{equation}\begin{split}\label{proba GM}
E_{\mathbb{P}} [\phi_{\theta}(L)\indic_{\{\tau>\theta\}} \,|\, \mathcal{G}^M_t]
&=E_{\mathbb{P}} [\phi_\theta(L)\indic_{\{ X^*_\theta > L\}} \,|\, \mathcal{F}_t \vee
\sigma(L)]\\&
={p_t(L)}^{-1} E_{\mathbb{P}^L}[\phi_\theta(L) p_{\theta}(L) \indic_{\{X^*_\theta > L\}} | \mathcal{F}_t \vee
\sigma(L)]\\
&={p_t(L)}^{-1} E_{\mathbb{P}^L} [\phi_\theta(x) p_{\theta}(x) \indic_{ \{X^*_\theta >
  x\}} | \mathcal{F}_t ]_{x=L}\\&
={p_t(L)}^{-1} E_{\mathbb{P}} [\phi_\theta(x) p_{\theta}(x) \indic_{ \{X^*_\theta >
  x\}} | \mathcal{F}_t ]_{x=L}.
  \end{split}
\end{equation}
\finproof

\rmk If $\mathcal{F}_{\theta}$ and  $\sigma(L)$  are
independent under  $ \mathbb{P}$, then $p_t(x)\equiv 1$, we obtain the  simpler formula \[E_{\mathbb{P}} [\phi_{\theta}(L)\indic_{\{\tau>\theta\}} \,|\, \mathcal{G}^M_t] = E_{\mathbb{P}} [\phi_\theta(x) \indic_{ \{X^*_\theta >
  x\}} | \mathcal{F}_t ]_{x=L} .\]

\begin{Pro} \label{pro:full}
We keep the notation of Section \ref{Sec:general} and define
$F_t^M(x):=p_t(x)\indic_{\{X_t^*>x\}}$. The value process of the
contingent claim $(C,G,Z)$ given the full information $(\G_t^M)_{t\geq 0}$ is
\begin{equation}\label{Equ:Man}
V_t^M=\indic_{\{\tau>t\}}\frac{\widetilde
V^M_t(L)}{p_t(L)}\end{equation} where
\begin{equation}\label{VMtilde}\widetilde V_t^M(L)={R_t}E_{\mathbb P}
\bigg[CR_T^{-1}F_T^M(x)+\int_t^TF_s^M(x)R_s^{-1}dG_s-
\int_t^TZ_sR_s^{-1}dF_s^M(x)\,\bigg|\,\mathcal F_t\bigg]_{x=L}.
\end{equation}
\end{Pro}
\bproof Using Proposition \ref{Propprobfull}, the first part of
(\ref{Equ:value process}) is given by
$$R_tE_{\mathbb{P}} \Big[C \indic_{\{\tau > T\}}R_T^{-1}| \mathcal{G}_t^M\Big]=
\frac{R_t}{p_t(L)} E_{\mathbb{P}}\big[ C R_T^{-1} p_{T}(x)
\indic_{\{ X^*_T >
  x\}} | \mathcal{F}_t \big]_{x=L}.$$
Let's see the third  term
$$R_t E_{\mathbb{P}} \Big[ Z_{\tau} R^{-1}_{\tau} \indic_{\{ t <\tau \leq  T\}}|
\mathcal{G}_t^M\Big].$$ We begin by assuming that $Z$ is a
stepwise $\mathbb F$-predictable process as in \cite{BR}, that is
$Z_u= \sum_{i=0}^n Z_i \indic_{t_i < u \leq t_{i+1}}$  for $t<u
\leq T$ where $t_0=t < \cdots < t_{n+1}=T$ and $Z_i$ is
$\mathcal{F}_{t_i}$-measurable for $i=0, \cdots, n$. We have
 \begin{eqnarray*}
&&E_{\mathbb{P}} \Big[ Z_{\tau} \indic_{\{ t <\tau \leq  T\}}|
\mathcal{G}_t^M\Big]  \,\,\,\,\\
&=&\sum_{i=0}^n   \left( \frac{1}{p_t(L)}  E_{\mathbb{P}^L}\Big[ Z_{i} p_{t_i}(L)  \indic_{\{ t_i <\tau\}}
|\mathcal{G}_{t}^M \Big]  -  \frac{1}{p_t(L)}  E_{\mathbb{P}^L}\Big[Z_{i} p_{t_{i+1}}(L) \indic_{\{ t_{i+1} <\tau\}}
| \mathcal{G}_t^M\Big] \right) \\
&=&\sum_{i=0}^n    \frac{1}{p_t(L)}  \left(  E_{\mathbb{P}}\Big[ Z_{i}
  p_{t_i}(x)  \indic_{ \{ X^*_{t_i}>x\}}
|\mathcal{F}_{t}\Big]  -    E_{\mathbb{P}}\Big[Z_{i}
p_{t_{i+1}}(x) \indic_{  \{ X^*_{t_{i+1}}>x\}}
| \mathcal{F}_t\Big] \right)_{x=L} \\
&= &  \frac{1}{p_t(L)} E_{\mathbb{P}} \left[ \sum_{i=0}^n Z_i  \left(
  p_{t_i}(x)  \indic_{ \{ X^*_{t_i}>x\}}
  -
p_{t_{i+1}}(x) \indic_{  \{ X^*_{t_{i+1}}>x\}}
 \right) | \mathcal{F}_t \right]_{x=L} .
\end{eqnarray*}
We define $F^M_t(x) =
p_{t}(x) \indic_{  \{ X^*_{t} > x\}}$. For $x$ fixed, $ \indic_{  \{
  X^*_{t} > x\}}$ is decreasing and right continuous, and according to \cite {Jac2},
$(p_{s}(x))_{s \geq 0}$ is an $(\mathbb{F}, \mathbb{P})$-martingale. Thus  $(F_t^M(x))_{t \geq 0 }$ is a nonnegative
$(\mathbb{F}, \mathbb{P})$-supermartingale, and   we may  deal
with its  right-continuous modification with finite left-hand limits.
Therefore
\begin{eqnarray*}
E_{\mathbb{P}} \Big[ Z_{\tau} \indic_{\{ t <\tau \leq  T\}}|
\mathcal{G}_t^M\Big] &=&   -\frac{1}{p_t(L)} E_{\mathbb{P}} \left[ \sum_{i=0}^n Z_i (F^M_{t_{i+1}}(x)-F^M_{t_{i}}(x)) | \mathcal{F}_t \right]_{x=L} \\
& = &  -\frac{1}{p_t(L)} E_{\mathbb{P}} \left[ \int_t^T  Z_u dF^M_{u}(x) | \mathcal{F}_t \right]_{x=L}.
\end{eqnarray*}
Finally, we get the third term of (\ref{Equ:value process}) by approximating $(Z_u  R^{-1}_u)_u$ by a suitable sequence of
 stepwise $\mathbb F$-predictable processes :
$$R_t E_{\mathbb{P}} \Big[ Z_{\tau} R^{-1}_{\tau} \indic_{\{ t <\tau \leq  T\}}|
\mathcal{G}_t^M\Big]=-\frac{R_t}{p_t(L)} E_{\mathbb{P}} \Big[  \int_t^T  Z_u  R^{-1}_u  dF^M_{u}(x) | \mathcal{F}_t \Big]_{x=L} .$$
The second term  of (\ref{Equ:value process}) can be decomposed in two parts as follows
\begin{eqnarray*}
 &&R_t E_{\mathbb{P}}  \Big[  \int_t^T
\indic_{\{\tau>u\}}R_u^{-1}dG_u|\, \mathcal{G}_{t}^M \Big] \, \, \, \, \\
&=& R_t
E_{\mathbb{P}}  \Big[ \indic_{\{\tau>T\}}\int_t^T
 R^{-1}_u    dG_u + \indic_{\{t < \tau \leq T\}}\int_t^{\tau}
 R^{-1}_u dG_u |\, \mathcal{G}_{t}^M  \Big]\\
&=&
\frac{R_t}{p_t(L)} E_{\mathbb{P}} \Big[p_{T}(x)  \indic_{\{ X^*_T >
  x\}} \int_t^T
 R^{-1}_u    dG_u   - \int_t^T \int_t^{u}
 R^{-1}_s dG_s      dF^M_{u}(x)  | \mathcal{F}_t \Big]_{x=L}.
\end{eqnarray*}
Putting the three terms all together leads to
\[V_t^M=\frac{R_t}{p_t(L)}E_{\mathbb P}
\bigg[F_T^M(x)\bigg(CR_T^{-1}+\int_t^TR_u^{-1}dG_u\bigg)
-\int_t^T\bigg(Z_sR_s^{-1}+\int_t^sR_u^{-1}dG_u\bigg)
dF_s^M(x)\,\bigg|\,\mathcal F_t\bigg]_{x=L}.\]
The equality \eqref{Equ:Man} then follows by an integration by part.
\finproof

\section{Progressive information}\label{sec:prog}
\subsection{Pricing with progressive enlargement of filtration}

The progressive information on $L$ corresponds to the standard
information modelling in the credit risk literature where an
investor observes the default event when it occurs. Recall that
$$\mathbb{G}= (\G _t)_{t \geq 0} \, \, \mbox{ with } \, \G_t=\F_t\vee \D_t, $$
where $\D_t=\D_{t+}^0$, $\D_t^0=\sigma(\tau\wedge t)$. The pricing
formula \eqref{Equ:value process} when $\mathcal H_t$ is $\G_t$ is
well known. We recall it briefly below and  we refer to
\cite{BR,BJR} for a proof.

Recall that the $\mathbb
G$-compensator of $\tau$ (under the probability $\proba$) is the
$\mathbb G$-predictable increasing process $\Lambda^{\mathbb G}$
such that the process $(\indic_{\{\tau\leq t\}}-\Lambda^{\mathbb G}_t,t\geq 0)$ is a
$(\mathbb G,\proba)$-martingale. The process $\Lambda^{\mathbb G}$
coincides on the set $\{t\leq \tau\}$ with an $\mathbb
F$-predictable process $\Lambda^{\mathbb F}$, called the $\mathbb
F$-compensator of $\tau$. We define $S_t:=\mathbb
P(\tau>t\,|\,\mathcal F_t)=\mathbb P(X_t^*>L\,|\,\mathcal F_t)$,
which is the Az\'ema supermartingale of $\tau$. The following
result is classical (see \cite{JeuYor, BR, EJY2000}).
\begin{proposition}\label{Pro:prog}
For any $\theta\ge t$ and any $\mathcal F_\theta$-measurable random variable $\phi_\theta$, one has
\begin{equation}E_{\mathbb{P}}[\phi_\theta\indic_{\{\tau>\theta\}}\,|\,\mathcal G_t]=
\indic_{\{\tau>t\}}\frac{E_{\mathbb{P}}[\phi_\theta S_\theta\,|\,\mathcal F_t]}{S_t}.\end{equation}
where $S_t:=\mathbb P(\tau>t\,|\,\mathcal F_t)$. The value process for an investors given the progressive information flow $\mathbb G$ is
\begin{equation}\label{prixprog}V_t=\indic_{\{\tau>t\}}\frac{R_t}{S_t}E_{\mathbb P}
\bigg[R_T^{-1}S_TC+\int_t^TR_u^{-1}S_udG_u-\int_t^TR_u^{-1}Z_udS_u\,\bigg|\,\mathcal F_t
\bigg].\end{equation}
\end{proposition}
\begin{Rem}It is interesting to note the similitude between
the case of manager (Proposition \ref{pro:full}) and the case of
investor (Proposition \ref{Pro:prog}).  Comparing the pricing
formulas \eqref{Equ:Man},\eqref{VMtilde} and \eqref{prixprog}, we
observe that $F^M$ plays a similar role in the full information
case as $S$ does in the progressive information case.
\end{Rem}

The pricing formula for delayed information flow is similar since $\mathbb G^D$ is the progressive enlargement of $\mathbb F^D$ with respect to $\tau$ and $\mathbb F^D$ is a sub-filtration of $\mathbb F$. The only difference is that $S_t$ and $R_t$ are not $\F_t^{D}$-measurable. 

\begin{proposition}\label{Pro:prog2}
For any $\theta\ge t$ and any $\mathcal F_\theta$-measurable random variable $\phi_\theta$, one has
\begin{equation}E_{\mathbb{P}}[\phi_\theta\indic_{\{\tau>\theta\}}\,|\,\mathcal G_t^{D}]=\indic_{\{\tau>t\}}\frac{E_{\mathbb{P}}[\phi_\theta S_\theta|\mathcal F_t^D]}{E_{\mathbb{P}}[S_t|\mathcal F_t^D]}\end{equation}
The value process for a delay-informed investors is
\begin{equation}\label{prixdelayed}V^D_t=\frac{\indic_{\{\tau>t\}}}{\esp[S_t|\F_t^{D}]}\,
\esp_{\mathbb{P}}\bigg[\frac{R_t}{R_T}S_T C+\int_t^T
\frac{R_t}{R_u}S_udG_u-\int_t^T\frac{R_t}{R_u}Z_udS_u\,\bigg|\,\F_t^{D}\bigg].
\end{equation}

\end{proposition}

\subsection{Intensity hypothesis}

In the reduced-form approach of credit risk modelling, the
standard hypothesis is the existence of the intensity of default
time $\tau$. We say that $\tau$ has an $\mathbb
F$-\emph{intensity} if its $\mathbb F$-compensator
$\Lambda^{\mathbb F}$ is absolutely continuous with respect to the
Lebesgue measure,  that is, there exists an $\mathbb{F}$-adapted
process $\lambda^{\mathbb F}$ (called the $\mathbb{F}$-{\it
intensity} of $\tau$ under $\mathbb{P}$) such that
$(\indic_{\{\tau\le t\}} - \int_0^{ t \wedge
\tau}{\lambda^{\mathbb F}_s} ds,t\geq 0)$ is a $(\mathbb{G},
\mathbb{P} )$-martingale. The intensity hypothesis implies that
$\tau$ avoids the $\mathbb{F}$-predictable stopping times and that
$\tau$ is $\mathbb{G}$ totally inaccessible.

Under the intensity hypothesis, the Doob-Meyer decomposition of
the supermartingale $S$ has the explicit form: the process
$(S_t+\int_0^tS_u\lambda_u^{\mathbb F}du, t\ge 0)$ is an $\mathbb
F$-martingale. The pricing formulae \eqref{prixprog} and
\eqref{prixdelayed} can be written as
\begin{gather}
V_t=\frac{\indic_{\{\tau>t\}}R_t}{S_t}E_{\mathbb P}
\bigg[R_T^{-1}S_TC+\int_t^TR_u^{-1}S_udG_u+
\int_t^TR_u^{-1}Z_uS_u\lambda_u^{\mathbb F}du\,\bigg|\,\mathcal
F_t
\bigg],\\
V^D_t=\frac{\indic_{\{\tau>t\}}}{\esp[S_t|\F_t^{D}]}\,
\esp_{\mathbb{P}}\bigg[\frac{R_t}{R_T}S_TC+\int_t^T
\frac{R_t}{R_u}S_udG_u+\int_t^T\frac{R_t}{R_u}Z_uS_u\lambda_u^{
\mathbb F}du\,\bigg|\,\F_t^{D}\bigg].
\end{gather}

Note that the intensity does not always exist. For example, in
the structural model where $L$ is deterministic, $\tau$ is a
$\mathbb F$ predictable stopping time. Hence its intensity does
not exist. It is in general a difficult problem  to determine the
existence of the intensity process (see \cite{GJZ}, \cite{GZ} for
a detailed discussion).

In contrast to the notion of intensity as above, the {\it default
intensity} in the credit analysis is often referred as the
instantaneous probability of default at time $t$ conditioned on
some filtration $(\mathcal H_t)_{t\geq 0}$:
\[\lambda_t=\lim_{\Delta t\rightarrow 0}\frac{1}{\Delta t}\proba(t<\tau\leq t+\Delta t |\mathcal H_t) \quad a.s.\]
Under Aven's conditions (see \cite{GJZ}, \cite{GZ}), the two  intensities coincide. But this
is not true in general. For example, in the classical structural
model, the default intensity equals to zero. However, the
intensity process does not exist in this case. The default
intensity when
$\mathcal{H}_t=\mathcal{F}_t^D$ has been studied in many
papers such as \cite{duffielando,C-DGH,jeanblancvalchev,GJZ}, the default  intensity is
strictly positive in the delayed information case.
We note that in the full information case where $\mathcal H_t=\G_t^M$, we encounter the same situation as in the structural model: the default intensity equals to zero since $L$ is $\G_t^M$-measurable.




\section{Noisy full information}\label{sec:noisy}
In this section, we consider the insider's information flow.
Recall that the insider has a perturbed information on the barrier $L$  which changes through time. We assume that the perturbation is given by an independent noise, and is getting clearer as time evolves. To be more precise,
the noised barrier is modeled by a process $(L_t=f(L,\epsilon_t))_{t\ge 0}$, where $f:\mathbb R^2\rightarrow\mathbb R$ is a given Borel measurable function, and $\epsilon$ is a process independent of $\mathcal F_\infty$. The information flow $\mathbb G^I=(\mathcal G_t^I)_{t\ge 0}$ of the insider is then given by
\[\mathcal G_t^I:=\mathcal F_t\vee\sigma(L_s,\,s\le t)\vee\D_t.\]

\subsection{Perturbed initial enlargement of filtration}\label{subsec:noisy}
 We firstly make precise the mathematical assumptions in this case.
We introduce an auxiliary filtration $\mathbb F^I=(\mathcal F_t^I)_{t\ge 0}$ defined as
\[\mathcal F_t^I:=\mathcal F_t\vee\sigma(L_s,\,s\le t).\]
Note that $\mathbb G^I$ is a progressive enlargement of $\mathbb F^I$ by the information on the default.
The filtration $\mathbb F^I$ has been studied in \cite{Cor} under Assumption \ref{hyp1}. It has nice properties similarly to the filtration $\mathbb G^M$. With the notation of Section \ref{Subsec:jacod}, assume that $\rho_t^I:=E_{\mathbb P}[\rho_t^M(L)|\mathcal F_t^I]$ satisfies $\int_0^\infty|\rho_t^I|dt<+\infty$ $\mathbb P$-a.s. Then the process $\widetilde B^I$ defined as
$\widetilde B^I_t:=B_t-\int_0^t\rho_s^Ids$ is an $(\mathbb F^I,\mathbb P)$-Brownian motion. Moreover, the Dol\'eans-Dade integral
\[Y_\cdot^I=\mathcal{E}  (  - \int_0^{\cdot} \rho^I_s d\widetilde{B}^I_s)\]
is a positive $(\mathbb F^I,\mathbb P)$-local martingale. We assume that $Y^I$ is an $(\mathbb F^I,\mathbb P)$-martingale and define the probability measure  $\mathbb Q^I$ by 
\[\mathrm{d}\mathbb Q^I=Y^I_t\mathrm{d}\mathbb Q\quad \text{ on $\mathcal F_t^I$}\]where $\mathbb Q$ is an equivalent probability of $\mathbb P$.
Then any $(\mathbb F,\mathbb Q)$-local martingale is an $(\mathbb F^I,\mathbb Q^I)$-local martingale. In particular, $B$ is an $(\mathbb F^I,\mathbb P^I)$-Brownian motion.

\subsection{Pricing with noisy information}
We now consider the pricing problem for the insider information flow $\mathbb{G}^I$.  We shall focus on the particular but useful case:
$$L_t=L+\epsilon_t, $$ where $\epsilon$ is a continuous process independent of $\F_\infty\vee\sigma(L)$ and is of
backwardly independent increments whose marginal has a density with respect to the Lebesgue measure (example in \cite{Cor} and \cite{moi}).
We say that a process $\epsilon$ has \emph{backwardly independent increments} if for all $0\le s\le t\le \theta$, the random variable $\epsilon_s-\epsilon_t$ is independent to $\epsilon_\theta$.
For example, if one takes $\epsilon_t= W_{g(T-t)}$ with $W$ an Brownian motion, and $g:[0,T] \rightarrow [0, + \infty) $  a strictly
increasing bounded function with $g(0)=0$, then $\epsilon$ is a process on $[0,T]$ which has backwardly independent increments. Another example with infinite horizon is $\epsilon_t=W_{g(\frac{1}{t+1})}$, where $g:[0,1] \rightarrow [0, + \infty) $  a strictly
increasing bounded function with $g(0)=0$.

To compute the pricing formula \eqref{Equ:value process} for the insider where $\mathcal H_t=\G_t^I$, our strategy is to combine the results in the two previous sections using the auxiliary filtration $\mathbb F^I$. More precisely, we present firstly in Proposition \ref{Propprobprogstrong} a result for the filtration $\mathbb F^I$ which is similar to the one in Proposition \ref{Propprobfull} for the filtration 
$\mathbb G^M$. We then use it to obtain the pricing formula in Theorem \ref{thm: pricing insider}. In fact, since $\mathbb G^I$ is the progressive enlargement of $\mathbb F^I$, applying \eqref{prixprog} leads to  the value process for insiders:
\begin{equation}\label{VIFI}
V_t^I=\frac{\indic_{\{\tau>t\}}R_t}{S_t^I}E_{\mathbb P}
\bigg[R_T^{-1}S_T^IC+\int_t^TR_u^{-1}S_u^IdG_u-
\int_t^TR_u^{-1}Z_udS_u^I\,\bigg|\,\mathcal F_t^I
\bigg]\end{equation}where $S_t^I:=E_{\mathbb P}[\indic_{\{\tau>t\}}|\mathcal F_t^I]$.
In the rest of the section, we aim to give a reformulation of \eqref{VIFI} as a conditional expectation with respect to the default-free filtration $\mathbb F$. It is interesting to remark that although the formula \eqref{Equ:probfullstr5.2} in Proposition \ref{Propprobprogstrong} seems to be complicated, the final result \eqref{Prix noisy} is given in a simple and coherent form similarly as for the full and progressive information. 

We assume Assumption \ref{hyp1} in the sequel, that is, the conditional probability law of $L$ given $\mathcal F_t$ has a density $p_t(\cdot)$ with respect to the unconditioned probability law of $L$. 

\begin{Pro}\label{Propprobprogstrong}
We assume Assumption \ref{hyp1}. Let $\epsilon$ be a continuous process, independent of $\F_\infty\vee\sigma(L)$, and with
backwardly independent increments such that the probability law of
$\epsilon_t$ has a density $q_t(\cdot)$ with respect to the Lebesgue measure.
For any $t\geq 0$, let $L_t=L+\epsilon_t$ and
$\mathcal{F}_t^I=\F_t\vee\sigma(L_s,s\leq t)$. Then, for any $\theta\ge t$ and
any positive $\mathcal F_\theta\otimes\mathcal B(\mathbb R)$-measurable
function $\phi_\theta(\cdot)$, one has
 {\small
\begin{equation}\label{Equ:probfullstr5.2}
 E_{\mathbb{P}} [\phi_\theta(L_\theta)\indic_{\{\tau>\theta\}} | \mathcal{F}^I_t]
=  \frac{ \iint
   E_{\mathbb{P}}[\phi_\theta(u+y) p_{\theta}(l) \indic_{\{X^*_\theta >
  l\}} | \mathcal{F}_t ]_{u=L_t}     q_{t}(L_t-l) \mu_{t,\theta}(dy)P^L(dl)
}{ \int_{\mathbb{R}} p_t(l)q_{t}(L_t-l) P^L(dl)    } 
\end{equation} }
where $P^L$ is the probability law of $L$, $\mu_{t,\theta}$ is the probability law of $\epsilon_\theta-\epsilon_t$.
For any $\F_\theta$-measurable $\phi_\theta$, one has
\[
E_{\mathbb{P}}[\phi_\theta\indic_{\{\tau>\theta\}} | \mathcal{F}^I_t]=\frac{ \int
   E_{\mathbb{P}} [\phi_\theta p_{\theta}(l) \indic_{\{X^*_\theta >
  l\}} | \mathcal{F}_t ] q_{t}(L_t-l) P^L(dl)    }{ \int_{\mathbb{R}} p_t(l)q_{t}(L_t-l) P^L(dl)    }.
\]
\end{Pro}

\bproof
Since $\epsilon$ has backwardly independent increment and is independent of $\mathcal{F}_{{\theta}}  \vee \sigma(L)$, one has
\begin{equation}\label{in proof}\begin{split}E_{\mathbb{P}} [\phi_\theta(L_\theta)\indic_{\{\tau>\theta\}} | \mathcal{F}^I_t]&=E_{\mathbb{P}}[\phi_\theta(L+\epsilon_\theta)\indic_{\{X_\theta^*>L\}} | \mathcal{F}\vee\sigma(L_t)\vee\sigma(\epsilon_s-\epsilon_t,\,s\le t)]\\
&=E_{\mathbb{P}}[\phi_\theta(L+\epsilon_\theta)\indic_{\{X_\theta^*>L\}} | \mathcal{F}\vee\sigma(L_t)]
.\end{split}
\end{equation}
By the independence of $\mathcal{F}_{{\theta}}  \vee \sigma(L)$ and $\epsilon$, we obtain
\[\begin{split}&\quad\;E_{\mathbb P} \left[\phi_\theta(L_\theta)\indic_{\{ \tau>\theta\}} | \mathcal{F}_t \vee
\sigma(L_t) \vee     \sigma(L) \right]\\&= E_{\mathbb{P}} \left[\phi_\theta(L_\theta) \indic_{\{X^*_{\theta}>L\}} | \mathcal{F}_t \vee
\sigma(\epsilon_t)  \vee   \sigma(L) \right]\\
&=\int_{\mathbb R} E_{\mathbb P}[
\phi_{\theta}(L_t+y)\indic_{\{X^*_{\theta}>L\}} | \mathcal{F}_t \vee
\sigma(\epsilon_t)  \vee   \sigma(L)]\mu_{t,\theta}(dy)\\
&=\int_{\mathbb R}
E_{\mathbb{P}} \left[\phi_\theta(L+z+y)\indic_{\{ X^*_{\theta}>L\}} | \mathcal{F}_t \vee
    \sigma(L) \right]_{z=\epsilon_t}\mu_{t,\theta}(dy)\\
    &=p_t(L)^{-1}\int_{\mathbb R} E_{\mathbb{P}} [\phi_{\theta}(x+y+z)p_{\theta}(x) \indic_{ \{X^*_\theta >
  x\}} \,|\, \mathcal{F}_t ]_{\begin{subarray}{c}x=L\\
  z=\epsilon_t
  \end{subarray}}\mu_{t,\theta}(dy), \end{split}\]
where the last equality comes from Proposition \ref{Propprobfull}. In the rest of the proof, we denote by
\[H_t(L,L_t):=p_t(L)^{-1}\int_{\mathbb R} E_{\mathbb{P}} [\phi_{\theta}(u+y)p_{\theta}(x) \indic_{ \{X^*_\theta >
  x\}} \,|\, \mathcal{F}_t ]_{\begin{subarray}{c}x=L\\
  u=L_t
  \end{subarray}}\mu_{t,\theta}(dy).\]
By definition and similar argument as for \eqref{in proof}, one has
\[\begin{split}E_{\mathbb P} [\phi_\theta(L_\theta)\indic_{\{\tau>\theta\}} | \mathcal{F}^I_t]
&= E_{\mathbb P}[H_t(L,L_t)|\mathcal F_t\vee\sigma(L_t)\vee
\sigma((\epsilon_t-\epsilon_s), s \leq t)]\\
&=E \left[ H_t(L,L_t)  | \mathcal{F}_t \vee
\sigma(L_t)  \right].
\end{split}\]
Let $P^L_t(dl)$ be the regular conditional probability of $L$ given
$\mathcal{F}_t $. Then for $U \in \mathcal{B}(\mathbb{R}^2)$,
$$ \mathbb{P} \left( (L,L_t) \in U | \mathcal{F}_t \right) = \int_{\mathbb{R}^2}
\indic_U(l,x) q_{t}(x-l) P^L_t(dl) dx$$
Therefore
\begin{equation}\label{cleprogstrong}
E \left[ H_t(L,L_t) | \mathcal{F}_t^I  \right]=\frac{ \int_{\mathbb{R}}
H_t(l,L_t) q_{t}(L_t-l) P^L_t(dl)    }{ \int_{\mathbb{R}} q_{t}(L_t-l) P^L_t(dl)
}.
\end{equation}
By the equality $P^L_t(dl)=p_t(l)P^L(dl)$,  we obtain the desired result. The second equality is obtained in a similar way.
\finproof

As a consequence of Proposition \ref{Propprobprogstrong}, the conditional expectation $E_{\mathbb P}[\indic_{\{\tau>t\}}|\mathcal F_t^I]$ can be written as $S_t^I(L_t)$, where $S_t^I(\cdot)$ is the $\mathcal F_t\otimes\mathcal B(\mathbb R)$-measurable function defined as
\begin{equation}S_t^I(x)=\frac{\int_{\mathbb R} \indic_{\{X_t^*>l\}}p_t(l)q_t(x-l)P^L(dl)}{
\int_{\mathbb R} p_t(l)q_t(x-l)P^L(dl)}.\end{equation}
In the following result, we compute  \eqref{VIFI} as $\mathbb F$-conditional expectations.



\begin{theoreme}\label{thm: pricing insider}
We keep the notations and assumptions of Proposition \ref{Propprobprogstrong} and recall that $\mathcal{G}_t^I=\F_t^I\vee\D_t$.
Then the value process for the noisy full information flow $\mathbb G^I$ is given by
\begin{equation}\label{Prix noisy}
{V_t^I}= \frac{\indic_{\{\tau>t\}}}{\int_{\R}F_t^M(l)q_t(L_t-l)P^L(dl)}\int\widetilde V^M_t(l)q_t(L_t-l)P^L(dl)
\end{equation}
where $\widetilde V^M$ and $F^M$ are defined in Proposition \ref{pro:full}.
\end{theoreme}

\bproof To obtain results with respect to $\F_t$,
we shall calculate respectively the three terms of \eqref{VIFI} using Proposition \ref{Propprobprogstrong}.
Let $N_{t}(x):=\int_{\R}\indic_{\{X_t^*>l\}}p_t(l)q_t(x-l)P^L(dl)=\int_{\R}F_t^M(l)q_t(x-l)P^L(dl)$.
Firstly,
\[E_{\mathbb{P}}[\frac{C}{R_T}\indic_{\{\tau>T\}}|\G_t^I]
=\indic_{\{\tau>t\}}\frac{E_{\mathbb{P}}[\frac{C}{R_T}\indic_{\{\tau>T\}}|\F_t^I]}{E_{\mathbb{P}}[\indic_{\{\tau>t\}}|\F_t^I]}=\frac{\indic_{\{\tau>t\}}}{N_t(L_t)}\int
E_{\mathbb{P}} \big[\frac{C}{R_T}F_T^M(l)|\F_t\big]q_t(L_t-l)P^L(dl)\] where
the second equality comes from Proposition \ref{Propprobprogstrong}.
Secondly, using the same argument,
\[\begin{split}&\quad
  E_{\mathbb{P}}[\int_t^T\indic_{\{\tau>\theta\}}\frac{dG_\theta}{R_\theta}|\G_t^I]=\int_t^T E_{\mathbb{P}} [\indic_{\{\tau>\theta\}}\frac{dG_\theta}{R_\theta}|\G_t^I]\\
&=\frac{\indic_{\{\tau>t\}}}{N_t(L_t)} \int \int_t^T E_{\mathbb{P}} [F_\theta^M(l)\frac{dG_\theta}{R_\theta}|\F_t] q_t(L_t- l)P^L(dl)\\
\end{split}\]
Thirdly, similar as in the proof of Proposition \ref{pro:full},
we assume  $Z_u= \sum_{i=0}^n Z_i \indic_{t_i < u \leq t_{i+1}}$  for $t<u \leq
T$ where $t_0=t < \cdots < t_{n+1}=T$ and $Z_i$ is
$\mathcal{F}_{t_i}$-measurable for $i=0, \cdots, n$. We have
 \begin{eqnarray*}
&& E_{\mathbb{P}} \Big[ Z_{\tau} \indic_{\{ t <\tau \leq  T\}}|
\mathcal{G}_t^I\Big]  \,\,\,\,\\
&=&\frac{\indic_{\{\tau>t\}}}{S_t^I} \sum_{i=0}^n
E_{\mathbb{P}}\Big[ Z_{i}   \indic_{\{ t_i <\tau\}} -Z_{i}  \indic_{\{ t_{i+1} <\tau\}}
| \mathcal{F}_t^I\Big] \\
&=&\frac{\indic_{\{\tau>t\}}}{N_t(L_t)}\int\sum_{i=0}^n
 \left(  E_{\mathbb{P}}\Big[ Z_{i}
  p_{t_i}(x)  \indic_{ \{ X^*_{t_i}>x\}}
|\mathcal{F}_{t}\Big]  -    E_{\mathbb{P}}\Big[Z_{i}
p_{t_{i+1}}(x) \indic_{  \{ X^*_{t_{i+1}}>x\}}
| \mathcal{F}_t\Big] \right) q_t(L_t-l)P^L(dl) \\
&= &  \frac{\indic_{\{\tau>t\}}}{N_t(L_t)}\sum_{i=0}^n  \int E_{\mathbb{P}} \left[ \sum_{i=0}^n Z_i  \left(
F_{t_i}^M(l)-F_{t_{i+1}}^M(l)
 \right) | \mathcal{F}_t \right]q_t(L_t-l)P^L(dl)
\\&=&  -\frac{\indic_{\{\tau>t\}}}{N_t(L_t)}\int E_{\mathbb{P}} \left[\int_t^T Z_u dF_{u}^M(l)
| \mathcal{F}_t\right] q_t(L_t-l)P^L(dl)
\end{eqnarray*}

We get the third term by approximating $(Z_u  R^{-1}_u)_u$ by a suitable sequence of
 stepwise $\mathbb F$-predictable processes :
$$E_{\mathbb{P}} \Big[ Z_{\tau} R^{-1}_{\tau} \indic_{\{ t <\tau \leq  T\}}|
\mathcal{G}_t^I\Big]= -\frac{\indic_{\{\tau>t\}}}{N_t(L_t)}\int E_{\mathbb{P}} \Big[\int_t^T \frac{Z_u}{R_u} dF_{u}^M(l)
 | \mathcal{F}_t\Big] q_t(L_t-l)P^L(dl).$$
We combine the three terms to complete the proof.
\finproof

\section{Risk-neutral pricing and numerical illustrations}\label{riskneutral}

\subsection{Pricing under different probabilities}

To evaluate a credit derivative, both the pricing filtration and the choice of risk-neutral probability
measures depend on the information level of the  market agent. 
In the previous sections, we have computed the pricing formula  \eqref{Equ:value process} for different information filtration under the same  historical probability measure. In the following, our objective is to take into account the pricing probabilities for each type of information.


We have made precise different pricing probabilities. First of all,  we assume that a pricing
probability $\Q$ is given with respect to the filtration $\mathbb{F}$ of the
fundamental process $X$. Usually, we choose  $\Q$ such that $X$ is an
$(\mathbb{F}, \Q)$ local martingale. Since we shall focus on the change of probability measures due to the different sources of informations and on its impact on the pricing of credit derivatives, we may assume, without loss of generality,  the historical probability $\mathbb{P}$ to be the benchmark pricing probability $\Q$ on $\mathbb{F}$. For the same reason, we will consider the same pricing probability for the filtration $\mathbb F$  and its progressive enlargement $\mathbb G$.\footnote{In general, a $(\mathbb{F}, \Q)$ local martingale is not necessarily a $(\mathbb{G}, \Q)$ local martingale except under (H) hypothesis. However, since all the filtrations we consider contains the progressive enlargement, we prefer to concentrate on the change of probabilities due to different sources of information and we keep the same pricing probability for $\mathbb F$ and $\mathbb G$.}
Given the pricing probability $\Q$ on $\mathbb F$ (and thus on $\mathbb G$), the pricing probability for the manager is $\Q^M$ where $\frac{d\Q^M}{d\mathbb{Q}}=Y^M(L)  $  with $Y^M(L)= \mathcal{E} (- \int_0^. \rho^M_s (L)(d B_s- \rho^M_s (L)ds))$ (see Subsection \ref{Subsec:jacod}) and  for the noisy full information  is
$\Q^I$ where  $\frac{d\Q^I}{d\mathbb{Q}}=Y^I $ with $Y^I= \mathcal{E}( -
\int_0^. \rho^I_s (dB_s- \rho^I_s ds))$ (see Subsection \ref{subsec:noisy}).
We also take $\Q$ as the pricing probability for the  delayed information
because the delayed information case is more
complicated : indeed, the notion of a $\mathbb{F}^D$ Brownian motion is a
widely open question that we do not want to investigate here and we assume that the pricing probability for the delayed case  is the same as  for the progressive information.

The following proposition gives the price of a credit derivative for the full
and the noisy information if we take into account not only the enlargement of
filtration but also the change of pricing probability due to this insiders'
information. Since we take $\mathbb{P}$ as the pricing measure, note that for the investors with progressive or delayed information, there is no change of pricing probability, so the results of Propositions \ref{Pro:prog} and \ref{Pro:prog2} still  hold.

\begin{Pro}\label{Propvaluefull6.1}
We assume Assumption \ref{hyp1}.\\
1) Define  $F^{\Q^M}_t(l) =\indic_{\{X_t^*>l\}}$.
Then the value process of a credit sensitive claim $(C,G,Z)$ for the manager's full information under the risk neutral  probability measure $\Q^M$  is given by
$$V_t^{\Q^M}= R_t \esp_{\mathbb{P}} \big[  C  R^{-1}_T  F^{\Q^M}_T(x)   +\int_t^T
 F^{\Q^M}_s(x) R^{-1}_s    dG_s -   \int_t^T  Z_s R^{-1}_s   dF^{\Q^M}_{s}(x) \, \,  | \mathcal{F}_t \big]_{x=L}.$$
2) Let  $\epsilon$ be a continuous process with
backwardly independent increments such that the probability law of $\epsilon_t$ has a density $q_t(\cdot)$ w.r.t. the Lebesgue measure. Then the value process for the insider's noisy full information under $\Q^I$ is given by
\begin{equation}
{V_t^{\Q^I}}= \frac{\indic_{\{\tau>t\}}}{\int_{\R}F_t^M(l)q_t(L_t-l)P^L(dl)}\int\widetilde V^{\Q^I}_t(l)q_t(L_t-l)P^L(dl)
\end{equation}
where
\begin{eqnarray*}
&&\widetilde V^{\Q^I}_t(l)= R_t \esp_{\mathbb{P}} \big[ C R_T^{-1} F_{t,T}^I(u,l)+ \int_t^T F_{t,\theta}^I(u,l) R^{-1}_\theta {dG_\theta}  - \int_t^T R^{-1}_\theta Z_\theta d F_{t,\theta}^I(u,l)   |\F_t\big]_{u=L_t} ,\\
&& F^I_{t,\theta}(u,l)=\mathcal{E}\Big(\int_t^\theta  \int \rho^I_\theta (u+y) \mu_{t,\theta}(dy) dB_u\Big)^{-1}F^M_\theta(l).
\end{eqnarray*}
\end{Pro}

To prove the second assertion of the above proposition, we need the following lemma which is an extension of Proposition \ref{Propprobprogstrong}. We give the proof of  Proposition \ref{Propvaluefull6.1} afterwards.

\begin{lemme}\label{Propprobprogstrongbis}
We keep the notations and assumptions of Proposition \ref{Propprobprogstrong}.
Then, for any $\theta\ge t$ and  any $\F_\theta$-measurable $\phi_\theta$, one has
\begin{equation*}\label{Equ:probfullstr}
\begin{split}
E_{\mathbb{P}} [ Y^I_\theta \phi_\theta \indic_{\{\tau>\theta\}} | \mathcal{F}^I_t]
= Y_t^I \frac{ \int  E_{\mathbb{P}} [\phi_{\theta}  F_{t,\theta}^I(u,l) \,|\, \mathcal{F}_t ]_{u=L_t}
q_{t}(L_t-l) P^L(dl)    }{ \int_{\mathbb{R}} p_t(l)q_{t}(L_t-l) P^L(dl)    },
\end{split}\end{equation*}
where $P^L$ is the probability law of $L$, $\mu_{t,\theta}$ is the probability law of $\epsilon_\theta-\epsilon_t$ and $F_{t,\theta}^I(u,l)$ is defined in Proposition \ref{Propvaluefull6.1}.
\end{lemme}
\bproof
First, let us recall, that $Y_T^I = \mathcal{E} ( \int_0^T \rho^I_u d B_u      )^{-1}$
and $\rho^I_t=E(\rho_t^M(L)| \mathcal{F}^I_t)= \frac{ \int \rho_t^M(l) q_t(L_t-l)P_t^L(dl)}{\int q_t(L_t-l)P_t^L(dl)}=\rho^I_t(L_t)$.
$(Y_t^I)_{t\geq 0}$ is an  $(\mathbb{F}^I, \mathbb{P})$ martingale.
Since $\epsilon$ has backwardly independent increment and is independent of $\mathcal{F}_{{\theta}}  \vee \sigma(L)$, one has
\[\begin{split}E_{\mathbb{P}} [\phi_\theta  Y^I_\theta \indic_{\{\tau>\theta\}} | \mathcal{F}^I_t]&= Y^I_t E_{\mathbb{P}}[\phi_\theta  \mathcal{E}(\int_t^\theta \rho^I_u(L_u) d B_u      )^{-1}
\indic_{\{X_\theta^*>L\}} | \mathcal{F}\vee\sigma(L_t)\vee\sigma(\epsilon_s-\epsilon_t,\,s\le t)]\\
&=Y^I_t E_{\mathbb{P}}[\phi_\theta  \mathcal{E}(\int_t^\theta \rho^I_u(L+\epsilon_u) dB_u)^{-1} \indic_{\{X_\theta^*>L\}} | \mathcal{F}\vee\sigma(L_t)]
.\end{split}
\]
By the independence of $\mathcal{F}_{{\theta}}  \vee \sigma(L)$ and $\epsilon$, we obtain
\[\begin{split}&\quad\;E_{\mathbb P} \left[\phi_\theta    \mathcal{E}(\int_t^\theta \rho^I_u(L+\epsilon_u) dB_u)^{-1}   \indic_{\{ \tau>\theta\}} | \mathcal{F}_t \vee
\sigma(L_t) \vee     \sigma(L) \right]\\
&= E_{\mathbb{P}} \left[\phi_\theta \mathcal{E}(\int_t^\theta \rho^I_u(L+\epsilon_u) dB_u)^{-1} \indic_{\{X^*_{\theta}>L\}} | \mathcal{F}_t \vee
\sigma(\epsilon_t)  \vee   \sigma(L) \right]\\
&=\int_{\mathbb R} E_{\mathbb P}[
\phi_{\theta}     \mathcal{E}(\int_t^\theta  \int \rho^I_u (L_t+y) \mu_{t,\theta}(dy) dB_u)^{-1}       \indic_{\{X^*_{\theta}>L\}} | \mathcal{F}_t \vee
\sigma(\epsilon_t)  \vee   \sigma(L)] \\
&=\int_{\mathbb R}
E_{\mathbb{P}} \left[\phi_\theta    \mathcal{E}(\int_t^\theta  \int \rho^I_u (L+z+y) \mu_{t,\theta}(dy) dB_u)^{-1}            \indic_{\{ X^*_{\theta}>L\}} | \mathcal{F}_t \vee
    \sigma(L) \right]_{z=\epsilon_t}\\
    &=p_t(L)^{-1}\int_{\mathbb R} E_{\mathbb{P}} [\phi_{\theta}p_{\theta}(x)    \mathcal{E}(\int_t^\theta  \int \rho^I_u (L+z+y) \mu_{t,\theta}(dy) dB_u)^{-1}                 \indic_{ \{X^*_\theta >
  x\}} \,|\, \mathcal{F}_t ]_{\begin{subarray}{c}x=L\\
  z=\epsilon_t
  \end{subarray}}, \end{split}\]
where the last equality comes from Proposition \ref{Propprobfull}. The rest of the proof is similar to the one of Proposition \ref{Propprobprogstrong} , with
\[H_t(L,L_t):=p_t(L)^{-1}\int_{\mathbb R} E_{\mathbb{P}} [\phi_{\theta}p_{\theta}(x)  \mathcal{E}(\int_t^\theta  \int \rho^I_u (u+y) \mu_{t,\theta}(dy) dB_u)^{-1}                 \indic_{ \{X^*_\theta >
  x\}} \,|\, \mathcal{F}_t ]_{\begin{subarray}{c}x=L\\
  u=L_t
  \end{subarray}}.\]
\finproof

\begin{proof} 
1) For the full manager, the  proof is  similar as the one of Proposition \ref{pro:full} : by noting that $\Q$ is chosen to be $\proba$, the probability measure $\Q^M$ coincides with $\proba^L$ defined in Section \ref{Subsec:jacod}.  Thus, the end of the proof of Proposition \ref{pro:full} still holds,   using $F_t^{\Q^M}=\indic_{\{X_t^*>l\}}$ instead of
$F_t^{M}$.\\
2) For the noisy information, $\frac{d\Q^I}{d\mathbb{P}}=Y_T^I$ with
$Y_T^I = \mathcal{E} ( \int_0^T \rho^I_u d B_u      )^{-1}$
and $\rho^I_t=\rho^I_t(L_t)$.
Let $N_{t}(x)=\int_{\R}\indic_{\{X_t^*>l\}}p_t(l)q_t(x-l)P^L(dl)=\int_{\R}F_t^M(l)q_t(L_t-l)P^L(dl)$ and $F^I_{t,\theta}(u,l)=\mathcal{E}(\int_t^\theta  \int \rho^I_\theta (u+y) \mu_{t,\theta}(dy) dB_u)^{-1}p_\theta(l)\indic_{ \{X^*_\theta >l\}}. $ For $T$, $l$ and  $u$ fixed,  $(F_{t,T}^I(u,l))_{0 \leq t \leq T }$ is a nonnegative
$(\mathbb{F}, \mathbb{P})$-supermartingale, and   we may  deal
with its  right-continuous modification with finite left-hand limits.
Firstly,
$$
\esp_{\mathbb{Q}^I}[\frac{C}{R_T}\indic_{\{\tau>T\}}|\G_t^I]
=\indic_{\{\tau>t\}}\frac{ \esp_{\mathbb{Q}^I}[\frac{C}{R_T}\indic_{\{\tau>T\}}| \mathcal{F}_t^I]}{ \mathbb{Q}^I [\tau>t |\mathcal{F}_t^I]}
=\indic_{\{\tau>t\}}\frac{\esp_{\mathbb{P}}[\frac{C}{R_T}Y_T^I\indic_{\{\tau>T\}}|\mathcal{F}_t^I]}{\esp_{\mathbb{P}}[\indic_{\{\tau>t\}} Y_t^I |\mathcal{F}_t^I]}$$
because on the event $\{\tau>t\}$,  $\frac{d\Q^I}{d\mathbb{P}}|_{\G_t^I}=\frac{d\Q^I}{d\mathbb{P}}|_{\F_t^I}=Y_t^I$.  Thus
$$\esp_{\mathbb{Q}^I}[\frac{C}{R_T}\indic_{\{\tau>T\}}|\G_t^I]
=\frac{\indic_{\{\tau>t \} }}{N_t(L_t)}\int\esp_{\mathbb{P}}\big[\frac{C}{R_T}F_{t,T}^I(u,l)|\F_t\big]_{u=L_t} q_t(L_t-l)P^L(dl)$$ where
the second equality comes from Lemma \ref{Propprobprogstrongbis}.
Secondly, using the same argument,
\begin{eqnarray*}
\esp_{\mathbb{Q}^I}[\int_t^T\indic_{\{\tau>\theta\}}\frac{dG_\theta}{R_\theta}|\G_t^I]&= &\indic_{\{\tau>t\}} \int_t^T \frac{\esp_{\mathbb{Q}^I} [\indic_{\{\tau>\theta\}}\frac{dG_\theta}{R_\theta}|\F_t^I]}{\mathbb{Q}^I [\tau>t|\F_t^I]}    \\
&=&\indic_{\{\tau>t\}}  \int_t^T  \frac{\esp_{\mathbb{P}}[   Y^I_\theta \indic_{\{\tau>\theta\}}\frac{dG_\theta}{R_\theta}|\F_t^I] }{\esp_{\mathbb{P}}[   Y^I_t \indic_{\{\tau>\theta\}} |\F_t^I] } \\
&=&\frac{\indic_{\{\tau>t\}}}{N_t(L_t)} \int \esp_{\mathbb{P}} \big[  \int_t^T F_{t,\theta}^I(u,l)\frac{dG_\theta}{R_\theta}|\F_t\big]_{u=L_t} q_t(L_t- l)P^L(dl)\\
\end{eqnarray*}
Thirdly,
we assume  $Z_u= \sum_{i=0}^n Z_i \indic_{t_i < u \leq t_{i+1}}$  for $t<u \leq
T$ where $t_0=t < \cdots < t_{n+1}=T$ and $Z_i$ is
$\mathcal{F}_{t_i}$-measurable for $i=0, \cdots, n$. We have
 \begin{eqnarray*}
&&E_{\mathbb{Q}^I} \Big[ Z_{\tau} \indic_{\{ t <\tau \leq  T\}}|
\mathcal{G}_t^I\Big]  \,\,\,\,\\
&&\indic_{\{\tau>t\}} \frac{E_{\mathbb{Q}^I} \Big[ Z_{\tau} \indic_{\{ t <\tau \leq  T\}}|
\mathcal{F}_t^I\Big]}{\mathbb{Q}^I [\tau>t|\F_t^I]}   \\
&=&\frac{\indic_{\{\tau>t\}}}{N_t(L_t)}\int\sum_{i=0}^n
 \left(  E_{\mathbb{P}}\Big[ Z_{i}
  Y^I_{t_i}  \indic_{ \{ X^*_{t_i}>x\}}
|\mathcal{F}_{t}\Big]  -    E_{\mathbb{P}}\Big[Z_{i}
Y^I_{t_{i+1}} \indic_{  \{ X^*_{t_{i+1}}>x\}}
| \mathcal{F}_t\Big] \right)_{u=L_t} q_t(L_t-l)P^L(dl) \\
&= &  \frac{\indic_{\{\tau>t\}}}{N_t(L_t)}\sum_{i=0}^n  \int E_{\mathbb{P}} \left[ \sum_{i=0}^n Z_i  \left(
F_{t_i,T}^I(u,l)-F_{t_{i+1},T}^I(u,l)
 \right) | \mathcal{F}_t \right]_{u=L_t} q_t(L_t-l)P^L(dl) \\
&=&  -\frac{\indic_{\{\tau>t\}}}{N_t(L_t)}\int E_{\mathbb{P}} \left[\int_t^T Z_s dF_{s,T}^I(u,l)
 | \mathcal{F}_t\right]_{u=L_t} q_t(L_t-l)P^L(dl)
\end{eqnarray*}
We conclude in the same way as in Proposition \ref{pro:full}.

\end{proof}

\subsection{Numerical examples}
We present numerical examples to illustrate the pricing formulas obtained previously.
We shall consider the following binomial model for the default barrier $L$.
\begin{ex}\label{L binomial model} (Binomial Model) \\
Let $L$ be a random variable taking two values   $l_i,l_s\in\R, \,l_i\leq l_s$ such that
$$\proba(L=l_i)=\alpha, \quad \proba(L=l_s)=1-\alpha\quad (0<\alpha<1).$$ Note that $L$ is independent of $(\F_t)_{t\geq 0}$.\\
\end{ex}

We suppose that the asset values process $X$ satisfies the Black Scholes model :
$$\frac{dX_t}{X_t}=\mu dt+\sigma dB_t, \quad t\geq 0.$$  
It is classical in this model to calculate conditional probabilities (\cite[Cor3.1.2]{BR}). In fact, 
for $t\geq 0$ and $h,l>0$,
\begin{eqnarray*}
\esp_{\proba}(\indic_{\{X^*_t>l\}}-\indic_{\{X^*_{t+h}>l\}}|\F_t)&=&\indic_{\{X^*_t>l\}}\left(\Phi\big(\frac{-Y^l_t-\nu h}{\sigma\sqrt h}\big)+e^{2\nu\sigma^{-2}Y_t}\Phi\big(\frac{-Y^l_t+\nu h}{\sigma\sqrt{h}}\big)\right)\end{eqnarray*}
where $\Phi$ is the standard Gaussian cumulative distribution function and  $$Y^l_t=\nu t+\sigma B_t +\ln \frac{X_0}{l}, \mbox{\,\,\, with\, \, \, } \nu=\mu-\frac 12 \sigma^2.$$
This formula will allow us to obtain explicit pricing results in the binomial default barrier model.

We give numerical comparisons of the value process of a defaultable bond  for different
information, in Example \ref{L binomial model} with the numerical values: $l_i=1, l_s=3, \alpha=\frac{1}{2}$.  We have fixed a very small constant delayed time, which makes the pricing results for the delayed information very close to the ones for the progressive information. 
We present in each figure two graphs, one being the dynamic price of a defaultable bond with zero recovery rate in the scenario of the firm value presented in the second graph.

In the scenario of Figure 1,  the manager has fixed the lower value for the  default threshold. So he estimates smaller default probability and thus higher price for the defaultable bond,  compared to the ones estimated by  other agents on the market.  We observe in addition that insider with noisy information has a better estimation of the price compared to the investors with progressive or delayed information. 
\begin{center}
\begin{figure}[h]\label{figure 1}
$$\begin{array}{ccc}
\includegraphics[width=7cm]{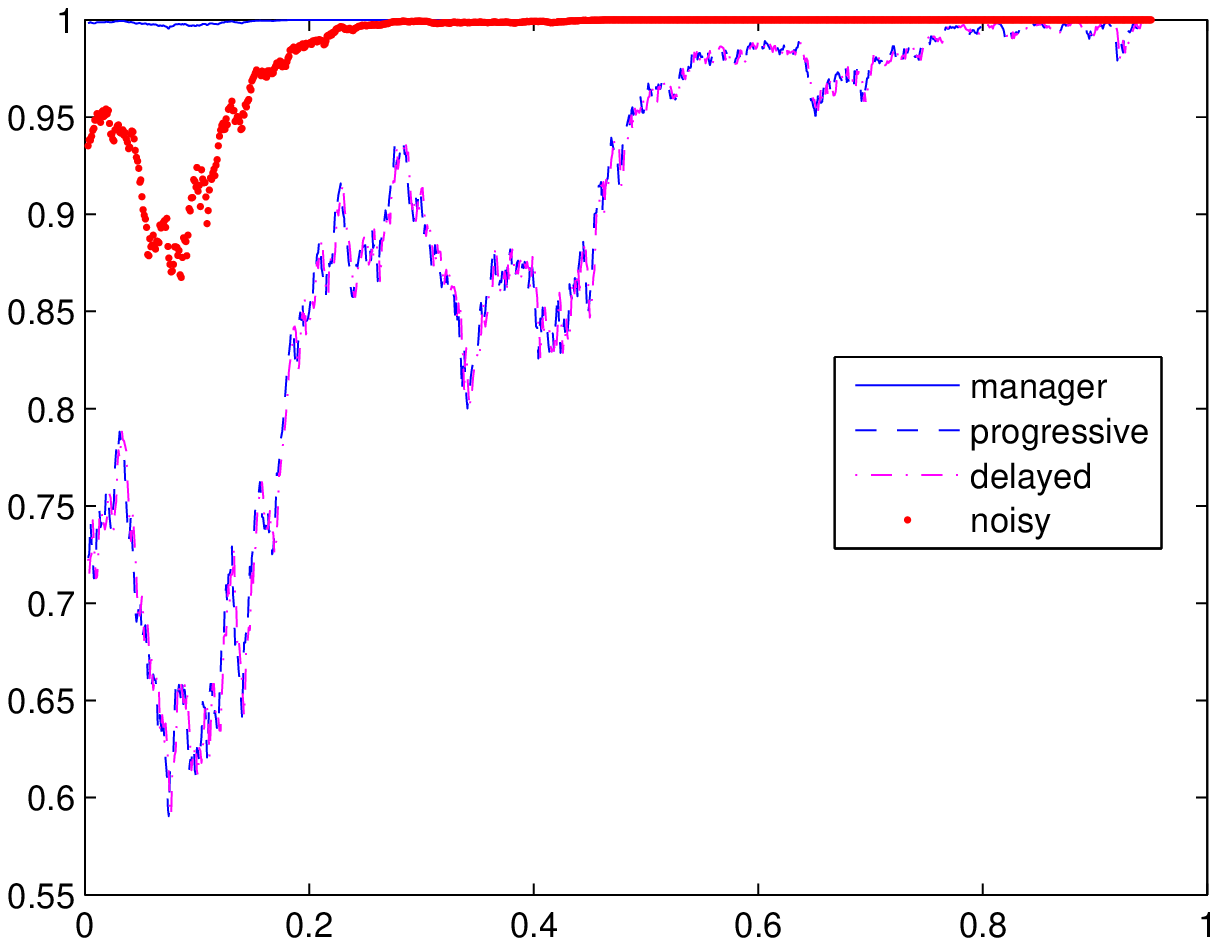} &
\includegraphics[width=6cm]{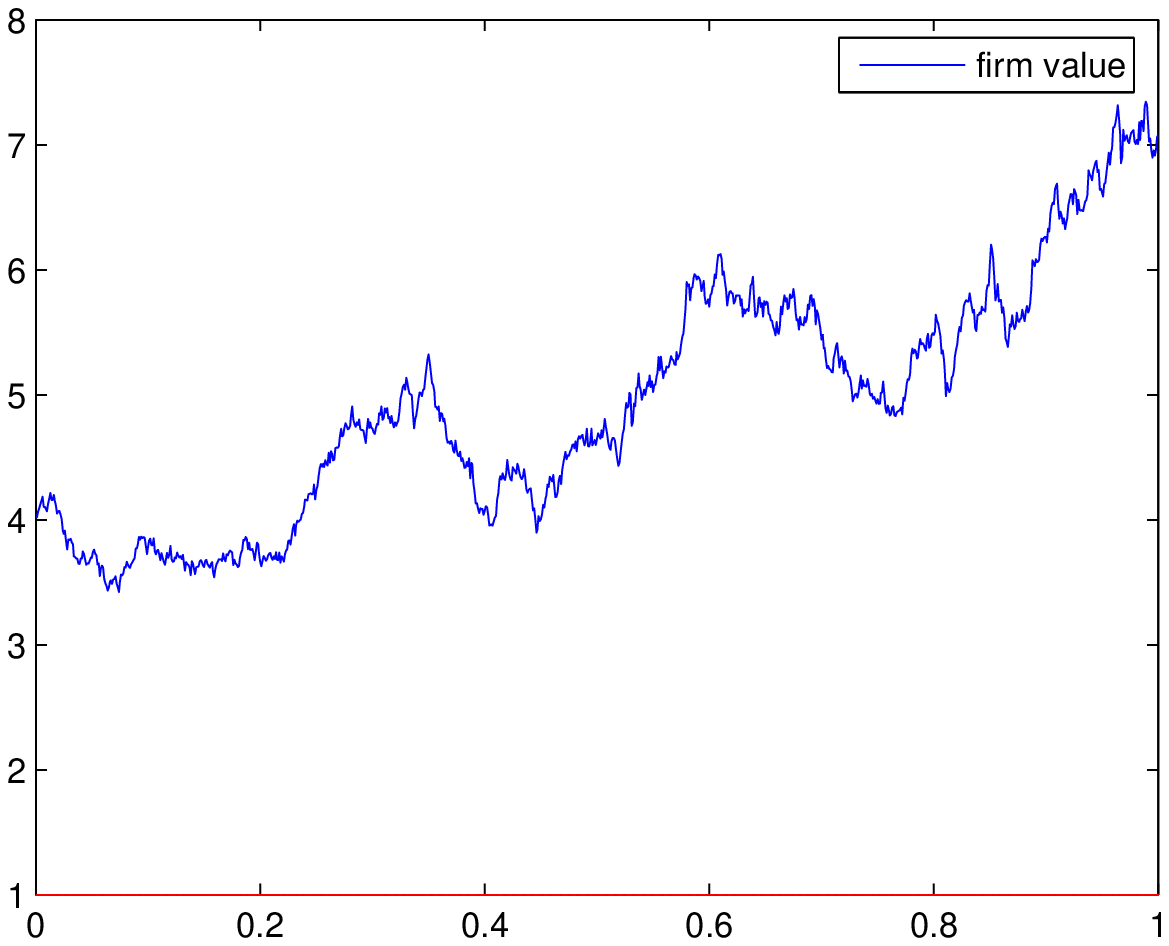} \\
\mbox{dynamic price of the defaultable bond} & \mbox{firm value }
\end{array}$$
\caption{$L=l_i$ }
\end{figure}
\end{center}

We  observe similar phenomena in  Figure 2:  the  manager has fixed the upper value for the default threshold and thus estimates higher probability of default  and smaller  price of the defaultable bond. Note that in the particular case where $L$ is constant ($l_i=l_s$), the price of the defaultable bond are the same, whatever the information we  consider.

\begin{center}
\begin{figure}[h]
$$\begin{array}{ccc}
\includegraphics[width=7cm]{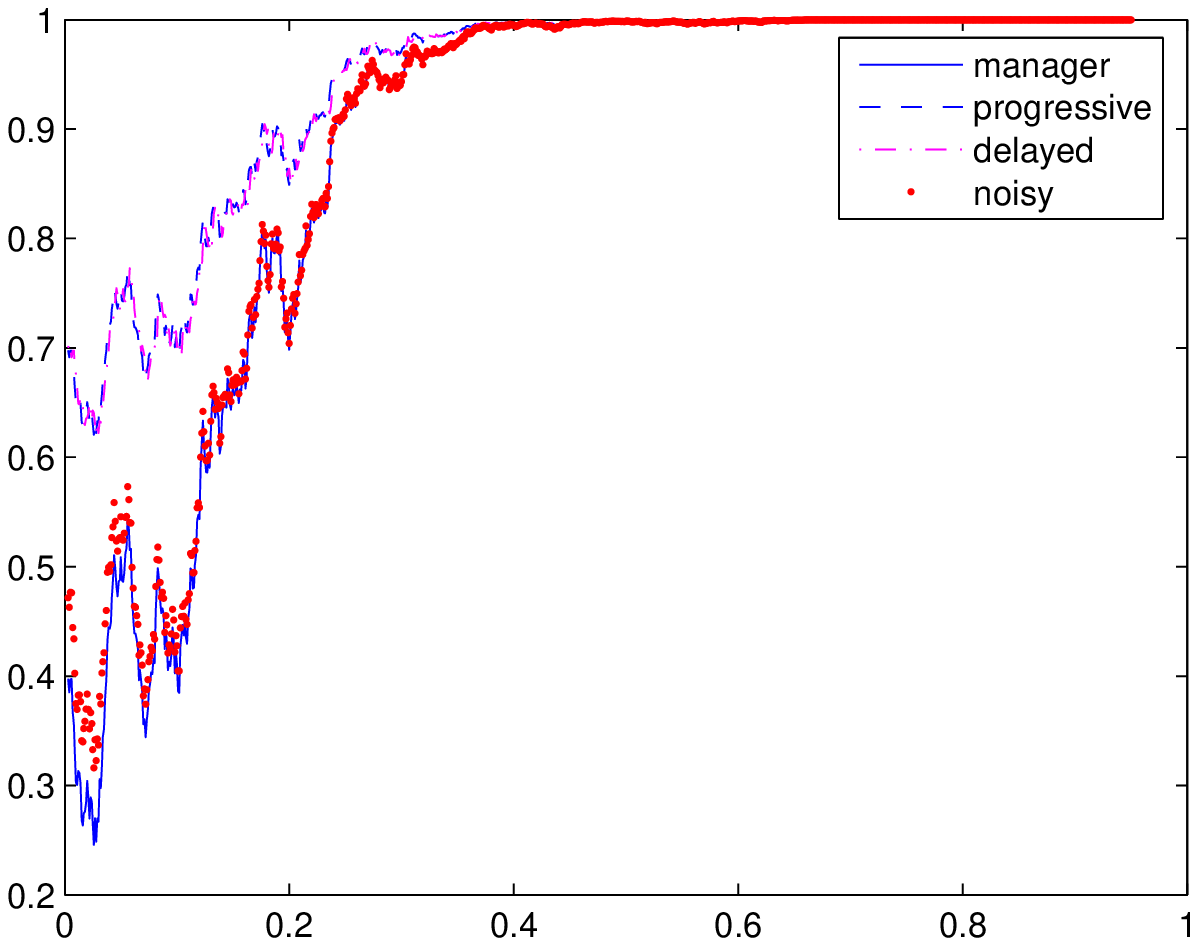} &
\includegraphics[width=7cm]{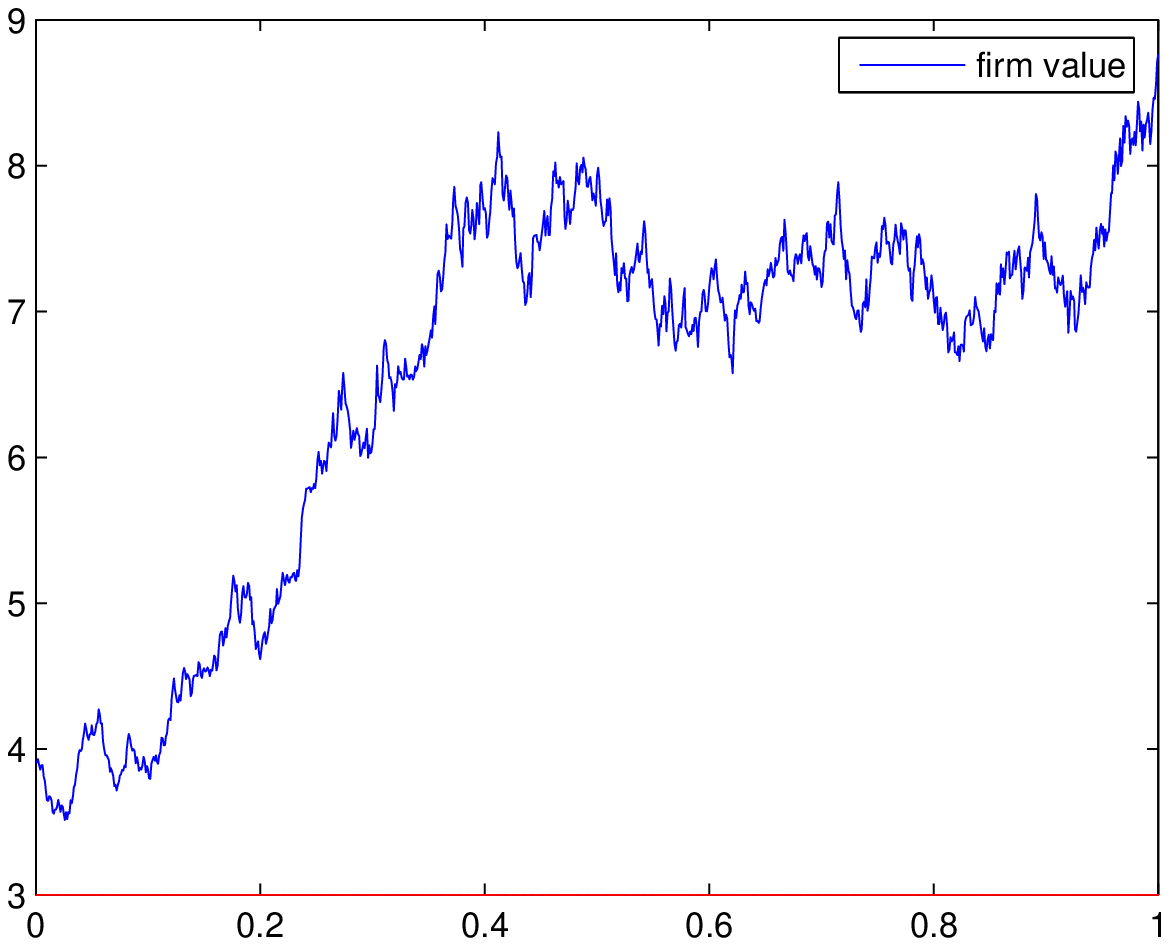} \\
\mbox{dynamic price of the defaultable bond } & \mbox{firm value }
\end{array}$$
\caption{$L=l_s$ }
\end{figure}
\end{center}

\vspace{-1.2cm}
\section{Conclusion}
We have modelled the different levels of default information by several types of enlargement of filtrations, leading also to different pricing probability measures. We have taken into account these two aspects in the pricing of credit derivatives and obtained  in all the cases coherent formulas given with respect to the ``default-free'' reference filtration. We have compared finally the pricing results by numerical illustrations.

\end{document}